\documentclass{PoS}
\usepackage{amssymb}
\usepackage{subfigure,epsfig,epstopdf}
\usepackage{bm}
\newcommand{\be}{\begin{equation}}

\newcommand{\ee}{\end{equation}}
\newcommand{\ba}{\begin{eqnarray}}
\newcommand{\ea}{\end{eqnarray}}
\usepackage{amsmath}

\title{Trapped phonons}

\ShortTitle{Trapped phonons}

\author{\speaker{Massimo Mannarelli}\\
        LNGS-INFN\\
        E-mail: \email{massimo@lngs.infn.it}}

\abstract{We analyze the effect of restricted geometries on the contribution of Nambu-Goldstone bosons (phonons)  to the shear viscosity, $\eta$, of a superfluid. For illustrative purpose we examine a simplified system consisting of a circular boundary of radius $R$, confining a two-dimensional rarefied gas of phonons. Considering the Maxwell-type conditions, we show  that phonons  that are not in equilibrium with the boundary and  that are not specularly reflected exert a shear stress on the boundary. In this case it is possible to define an effective (ballistic) shear viscosity coefficient  $\eta \propto \rho_{\rm ph} \chi R$, where $\rho_{\rm ph}$ is the density of phonons and $\chi$ is a parameter which characterizes the type of scattering at the boundary.   For an optically trapped superfluid our results corroborate the findings of Refs. \cite{Mannarelli:2012su, Mannarelli:2012eg}, which imply that at very low temperature the shear viscosity correlates with the size of the optical trap and decreases with decreasing temperature.} 

\FullConference{Xth Quark Confinement and the Hadron Spectrum,\\
		October 8-12, 2012\\
		TUM Campus Garching, Munich, Germany}

\begin{document}

\section{Introduction}
The  transport coefficients of a fluid depend on the  underlying microscopic dynamics  and on the boundary conditions imposed. The effect of the boundary conditions on the transport properties is seldom considered, but might nevertheless be important in particular circumstances. Actually a complete knowledge of the microscopic dynamics of a system should  include a detailed description of the boundary and  of the interactions of the particles  with the boundary. 

\begin{figure}[b]
\begin{center}
\includegraphics[width=12.cm]{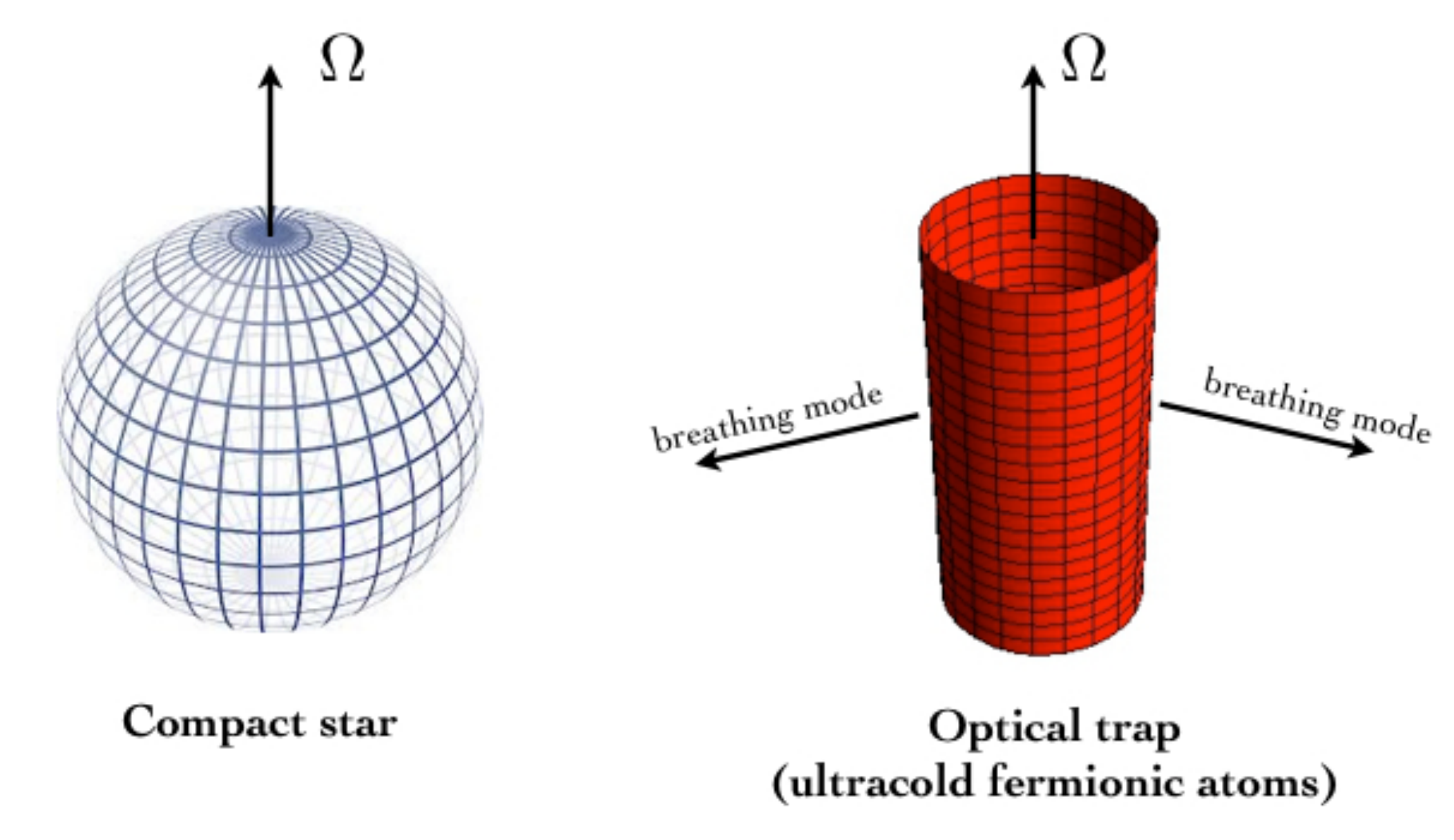}
%\subfigure{\includegraphics[width=5.cm]{compact-star.jpg}}\,\,\,\,\,\, \subfigure{\includegraphics[width=7.cm]{optical-trap.jpg}} 
\end{center}
\caption{Two pictorial examples of trapped geometries of physical interest. Left panel: Spinning compact star. Right panel: optical trap. In this case the trap can be put in rotation and/or it can be radially expanded. \label{fig:traps}}
\end{figure}

\begin{figure}[t]
\begin{center}
\includegraphics[width=7.5cm]{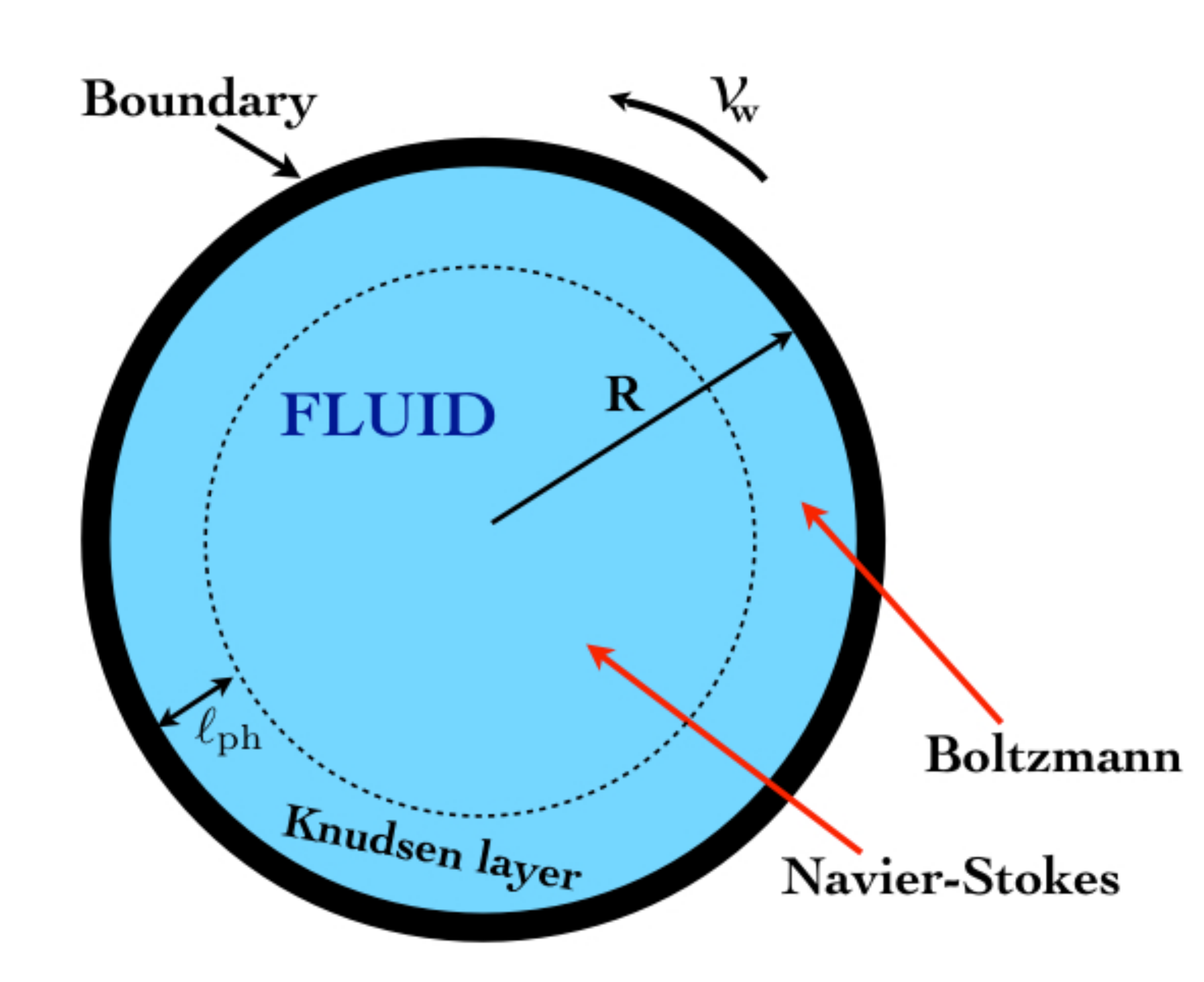}
\end{center}
\caption{Schematic representation of a two-dimensional trapped fluid. The boundary rotates with velocity $v_w = \Omega\, R$. Shown is the Knudsen layer,   where the Boltzmann equation must be used. The innermost  part of the fluid can be described by the usual Navier-Stokes equations.   \label{fig:circular-boundary}}
\end{figure}

Here we shall focus on a system consisting of Nambu-Goldstone bosons (phonons) with a \emph{simple boundary}, that is a boundary which does not allow  particle  transfer. Two different geometries are pictorially shown in Fig.\ref{fig:traps}, which correspond to a  spherically symmetric trap (left panel), relevant for the superfluid region of a compact star, and  a cylindrically symmetric trap (right panel), relevant for optically trapped ultracold atoms. In both cases we are interested to the evaluation of the drag force between the fluid and the boundary when the boundary rotates, with frequency $\Omega$, or it radially expands. However, for illustrative purposes we shall consider the simplified  two-dimensional system depicted in Fig.\ref{fig:circular-boundary}, corresponding to a circular rotating boundary of radius $R$, which can be thought as obtained by cutting  the rotating configurations of Fig.\ref{fig:traps} with a plane orthogonal to the $\Omega$-axis.

The effect of the boundary on the fluid dynamics is important in the so called Knudsen layer, which corresponds to a layer close to the boundary of width proportional to the mean free-path $\ell_{\rm ph}$. Any particle moving in this layer  is ballistic, meaning that it  has a probability of hitting the boundary  equal or larger than the probability of hitting a particle. The fluid in this region cannot be described by hydrodynamics; the corresponding Boltzmann equation with the proper boundary conditions should instead be used. In cases where $\ell_{\rm ph} \ll R$, the Knudsen layer can be neglected, and imposing usual (no-slip) boundary  conditions to the Navier-Stokes equations effectively takes into account the Knudsen layer.

However,  the finite extension of the trap does play an important role if the Knudsen number, $K_n = \ell_{\rm ph}/R$, is sizable: For $K_n \lesssim 0.1$  (say) the Navier-Stokes equation are still valid, but  the transport coefficients must be amended with  terms proportional to powers of $K_n$, which  depend as well on the trap geometry. For  $K_n \gg 1$ the system is ballistic and no hydrodynamic description can be used. For $0.1 \lesssim K_n \lesssim 1$ interparticle interactions and interactions with the boundary should be considered on an equal footing.

\section{Boundary description}
\begin{figure}[t]
\begin{center}
\includegraphics[width=12.cm]{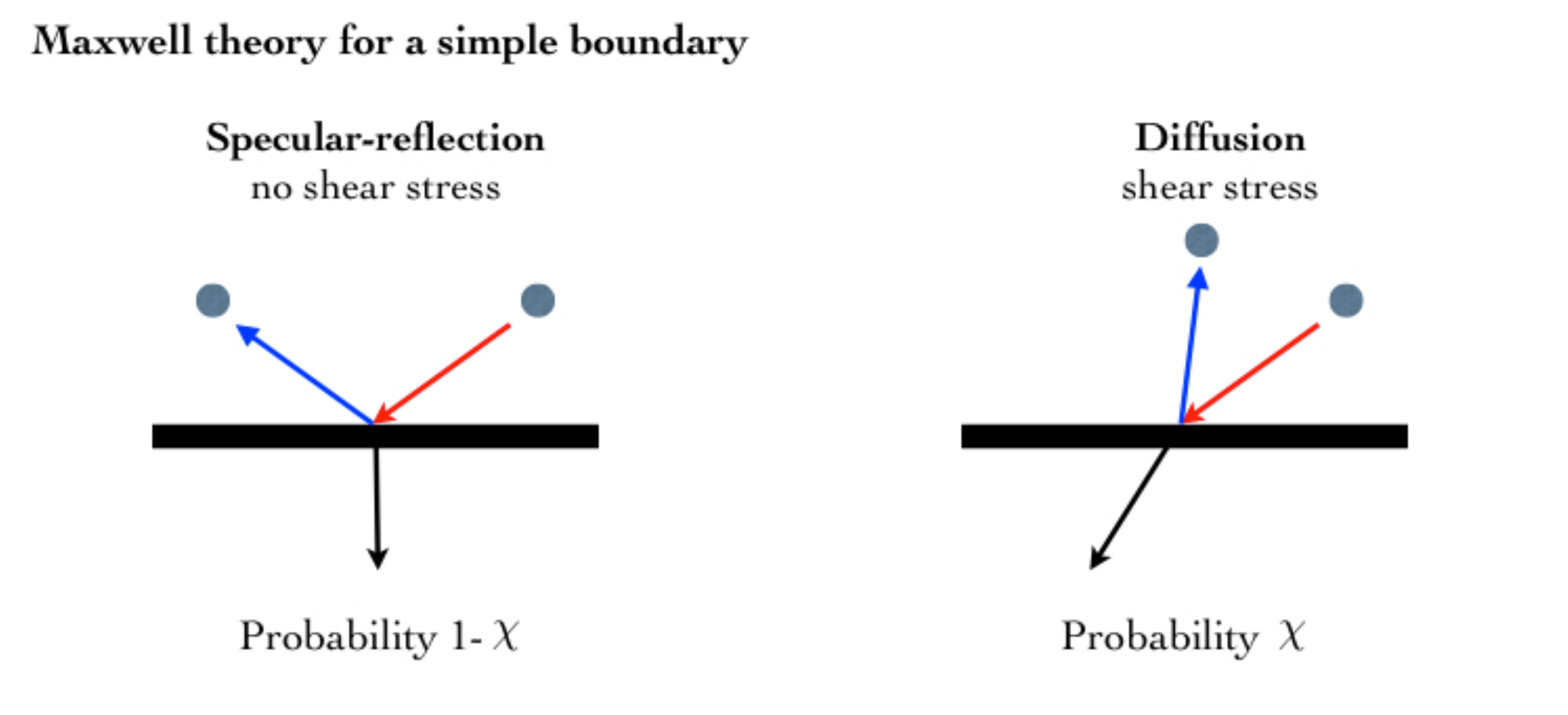}
\end{center}
\caption{Schematic representation of the Maxwell theory of a simple boundary. The red arrow corresponds to the momentum of the impinging particle; the blue arrow corresponds to the momentum of the scattered particle;  the black arrow corresponds to the momentum transferred to the boundary. Specular-reflection does not produce a shear stress on the boundary but a pressure orthogonal to the surface. Diffusion does produce a shear stress, because the  transferred momentum to the boundary is not orthogonal to the contact surface.  \label{fig:maxwell}}
\end{figure}

In the ballistic regime, $K_n \gg 1$, the distribution function of the particles can only change by interactions with  the boundary. In general,  the presence of a boundary changes the distribution function of the particles because the particle which have collided with the boundary will keep memory of it as far as they are in the Knudsen layer~\cite{Cercignani, Sone, Struchtrup}.
A simple boundary is equivalent to a wall that specular reflects or diffuses the impinging particles, and in this case  it is possible to use the  Maxwell-type boundary conditions, meaning that   the distribution function of the particles at the wall is given by
\be
\bar f=\left\{\begin{array}{lc} \chi f_w + (1-\chi) f(\bm x, \bm \xi(1 - 2(\bm \xi - \bm v_w)\cdot \bm n), t) &\text{for   } (\bm \xi - \bm v_w)\cdot \bm n>0 \\ f (\bm x, \bm \xi, t) &\text{for   } (\bm \xi - \bm v_w)\cdot \bm n<0 \end{array}\right.\,,\label{eq:barf}
\ee
where $f_w$ is the distribution function of the particles diffused by  the wall, $ \bm v_w \ll 1$ is the wall velocity, $\bm n$ is the unit normal vector to the wall, pointed to the fluid, and $\bm \xi = \bm p/p_0$ is the velocity of the particle. We shall further simplify the analysis assuming that $f_w$ is thermal. The \emph{accommodation coefficient}, $\chi$, describes the type of scattering at the boundary (schematically described in Fig.\ref{fig:maxwell});
the impinging particle is specularly reflected with probability $1 -\chi$ and is diffused with probability $\chi$. The distribution function $f$ describes the particles that hit the wall at the time $t$ and that are specularly reflected. Since we are concerned with a ballistic  fluid,  it corresponds to the distribution function of  the wall hit in a previous collision at the time $t_0$, that is    $f (\bm x, \bm \xi, t)  = f_w (\bm x- \bm \xi (t-t_0), \bm \xi, t_0) $, where $t_0 < t$. Clearly, if $\chi = 0$  phonons do not thermalize with the  wall and $f (\bm x, \bm \xi, t)$ corresponds to the initial  distribution function.   

Given the particle distribution function it is possible to determine  the shear stress on the wall from the off-diagonal components of the stress tensor 
\be
\tau_{ij}^C = \int (p_i - p v_{w,i}) (p_j - p v_{w,j}) \bar f (\bm x, \bm \xi, t) \frac{d \bm p}{(2 \pi)^2 E} \,,
\label{eq:tau}\ee
where $i,j$ are the cartesian coordinates. For the circular symmetry considered here, it is more appropriate the use polar coordinates $(r,\phi)$  and the corresponding stress tensor is obtained by  $\tau^P = U(\phi) \tau^C U(\phi)^T $, where  $U(\phi)$ is the rotation matrix. Upon plugging Eq.\eqref{eq:barf} in Eq.\eqref{eq:tau}, one readily obtains that the off-diagonal components vanish for $\chi=0$, consistent with the fact that specular-reflected particles do not produce shear stress on the boundary.  The shear stress  also vanishes if $f_w (\bm x- \bm \xi (t-t_0), \bm \xi, t_0) =  f_w (\bm x, \bm \xi, t)$; the reason is that in this condition the system is stationary and the  wall cannot be subject to neither force nor moment of force \cite{Sone}. 

Therefore, for the closed geometry considered, in order to  have a shear stress on the wall there must exist an external agent that perturbs the boundary. In the present case we shall assume that the boundary is put in rotation with a frequency $\Omega$. The perturbation starts  at a certain  time $t_i$ such that $t_0<t_i < t$ and the external agency maintains the system in rotation at the same frequency $\Omega$. The  impinging particles have  memory of the distribution function before the circle was put in rotation and will try to catch up with the wall. This produces a shear stress (force per unit length) on the wall given by
\be
\tau_{r \phi}(t) =-C\, \rho_{\rm ph} \Omega R \chi \,,
\ee
where $C$ is some dimensionless number  and $\rho_{\rm ph}$ is the two-dimensional phonon density
%\be \rho_{\rm ph} = \frac{2 \pi^2 T^4}{45 c_s^5}\,,\ee 
 of the normal fluid component. After a certain time, depending on $\chi$, $R$ and $c_s$,  the system will equilibrate and the phonons will ``corotate" with the boundary  not exerting any shear stress on it. The above expression of the stress tensor
allows us to define the ballistic (effective) shear viscosity coefficient for the non-equilibrated system given by 
\be
\eta_{\rm ball}^{\rm 2D} = C \,  \rho_{\rm ph}\, \chi\, R\,,
\label{eta-ball}
\ee
which should be rather general, meaning that it should be the valid also for the two geometries in Fig.\ref{fig:traps}, but  with a different expression of the constant $C$. We shall study this issue elsewhere. This expression has been used  in the study of the oscillations of compact stars in \cite{Andersson:2010sh, Manuel:2012rd}, for the description of the oscillations of objects immersed in $^4$He, see \emph{e.g.} \cite{Zadorozhko},  and for the breathing mode of ultracold fermionic atoms (discussed below) in \cite{Mannarelli:2012su, Mannarelli:2012eg}.

\section{Application to ultracold fermionic atoms}
The attractive interaction   between certain types of  fermionic  atoms (like $^6{\rm Li}$) 
prepared in two  hyperfine states can be varied at will using a magnetic-field Feshbach resonance.
These atoms can be confined in a cylindrically symmetric optical trap (schematically shown in the right panel of Fig.\ref{fig:traps}) and the system can be described as a core comprising phonons and gapped fermions and a corona of unpaired fermions, for a review see \cite{Giorgini:2008zz}. The extension of the corona  decreases with decreasing temperature and at $T\simeq 0.1 T_F$,   where $T_F$ is the Fermi temperature of the trapped system, the largest part of the system is superfluid.  The core of the system ceases to be superfluid for $T > T_c^{\rm trap} \simeq 0.21 T_F$ \cite{Haussmann:2008}.

 With decreasing temperature the entropy per particle steeply decreases and is not incompatible with the hypothesis that it receives a sizable contributions by phonons. Present experimental results do not allow to infer whether at the reached temperature the system is dominated by phonons~\cite{Ku}, however there is little doubt that at sufficiently low temperature phonons will dominate. 

The  phonon dispersion law (neglecting $O \left(k/k_F\right)^7$ terms) can be written as 
\be
\label{eq:disp-law}
E_k = c_s k (1 + \tilde\gamma\, (k/k_F)^2 + \tilde \delta\, (k/k_F)^4 )   \ ,
\ee 
where $\tilde\gamma  \simeq  0.18$  (see \cite{Rupak:2008xq, Salasnich, valle}  and the discussion in \cite{Mannarelli:2012eg}), and  a numerical estimate of $\tilde\delta$ was obtained in \cite{Mannarelli:2012su, Mannarelli:2012eg}  from a fit of the experimental values of the shear viscosity to entropy ratio. Indeed the contribution of binary (hereafter 4ph)  and $1  \leftrightarrow 2$  (hereafter 3ph) processes to the shear viscosity depends on the phonon dispersion law and in 
Refs. \cite{Mannarelli:2012su, Mannarelli:2012eg} the $\eta_{\rm 3ph}$ contribution was evaluated with the dispersion law in Eq. \eqref{eq:disp-law}. When compared with the experimental data obtained by the Duke group \cite{kinast1, kinast3}, one obtains $\tilde\delta = -(0.02 \div 0.04)$.

\begin{figure}[t]
\begin{center}
\includegraphics[width=8.cm]{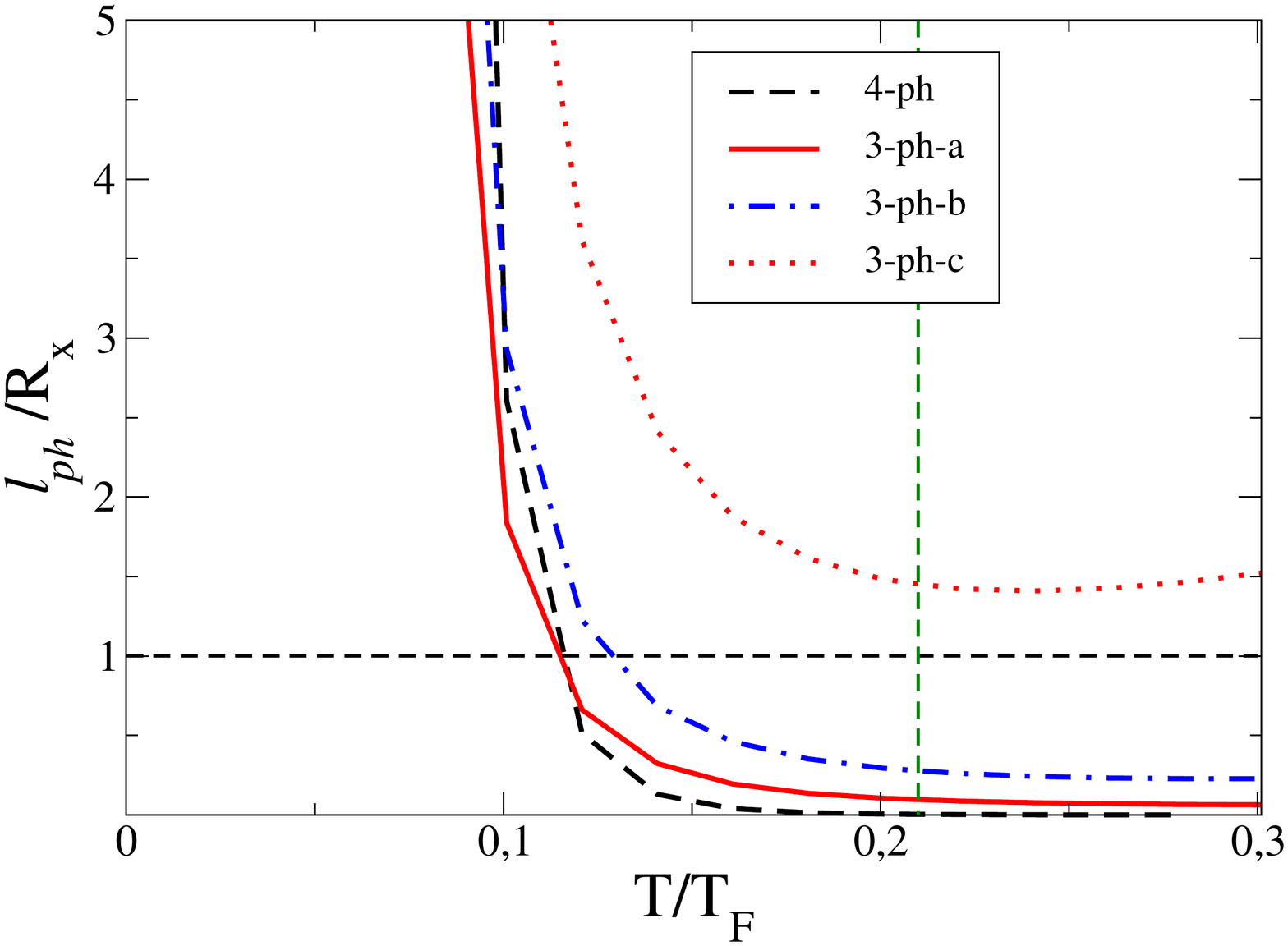}
\end{center}
\caption{Mean free-path of phonons in unit of $R_x$  as a function of $T/T_F$,  for  optically trapped superfluid fermionic atoms. The black dashed line corresponds to the mean free path associated to the 4ph process. The solid red line (named 3ph-a) corresponds to the  3ph process with $\tilde\delta\simeq 0.02$, the  dot-dashed blue line (named 3ph-b) corresponds to the 3ph process with $\tilde\delta\simeq 0.03$ and the dotted red line (named 3ph-c) corresponds to the  3ph process with $\tilde\delta\simeq 0.04$.  The horizontal black dashed line corresponds to  $K_n=1$. The hydrodynamic description is valid for $K_n \ll 1$. The vertical dashed green  line approximately corresponds to the transition temperature between the normal phase and the superfluid phase \cite{Haussmann:2008}.
Adapted from \cite{Mannarelli:2012eg}. \label{fig:free-path}}
\end{figure}
The contribution of the ballistic shear viscosity must be included in the description of the system because,  as  realized  in \cite{Mannarelli:2012su, Mannarelli:2012eg}   and shown in Fig.\ref{fig:free-path}, the mean free-path of phonons diverges at sufficiently low temperature.   In this case the typical length scale  appearing in Eq.\eqref{eta-ball} is the smallest radial extension of  the harmonic trap of \cite{kinast1, kinast3}, that is $R_x \simeq 7 \mu m$.

In principle, for any value of the Knudsen number  the shear viscosity can be obtained  plugging in Eq.\eqref{eq:tau} the distribution function solution of   the Boltzmann equation where  both  collisions of phonons with the boundary and the collisions among phonons are taken into account.  Solving this problem is complicated even for ideal gases \cite{Cercignani, Sone, Struchtrup}. However, since  we know the behavior of the relaxation time $\tau_b$  (corresponding to the time between two collisions with the boundary), as well as the relaxation time  $\tau_{\rm ph}$    (corresponding to the relaxation time of 3ph or 4ph processes), it is  possible to employ the same reasoning at the basis of the Matthiessen's rule \cite{Ashcroft-Mermin}  to obtain an approximate expression of $\eta$  at any temperature. Assuming that the two above-mentioned collisions are not correlated, one can   interpolate between the values of the shear viscosity coefficient in the ballistic and in the hydrodynamic regimes. For this purpose one can define an effective relaxation time
\be \tau^{-1}_{\rm eff} = \tau^{-1}_b + \tau^{-1}_{\rm ph}
\,,\ee
incorporating the effects of interparticle collisions and of the collisions with the boundary. 
Since the shear viscosity coefficient is proportional to the collision time,  we define the total effective shear viscosity as
\be
\eta_{\rm eff} = \left( \eta_{\rm  3ph}^{-1} + \eta_{\rm ball}^{-1}\right)^{-1} \,,
\label{eta_eff}
\ee
where $\eta_{\rm ball}$ is the three-dimensional analogous of Eq.\eqref{eta-ball} and $\eta_{\rm 3ph}$ is the 3ph shear viscosity \cite{Mannarelli:2012su, Mannarelli:2012eg}. 
In principle, the contribution of the 4ph collisions should be considered as well, but so far it has  not been determined  with the full expression of the dispersion law given in Eq.\eqref{eq:disp-law} and therefore it cannot be consistently added to Eq.\eqref{eta_eff}. 

\begin{figure}[t]
\begin{center}
\includegraphics[width=8.cm]{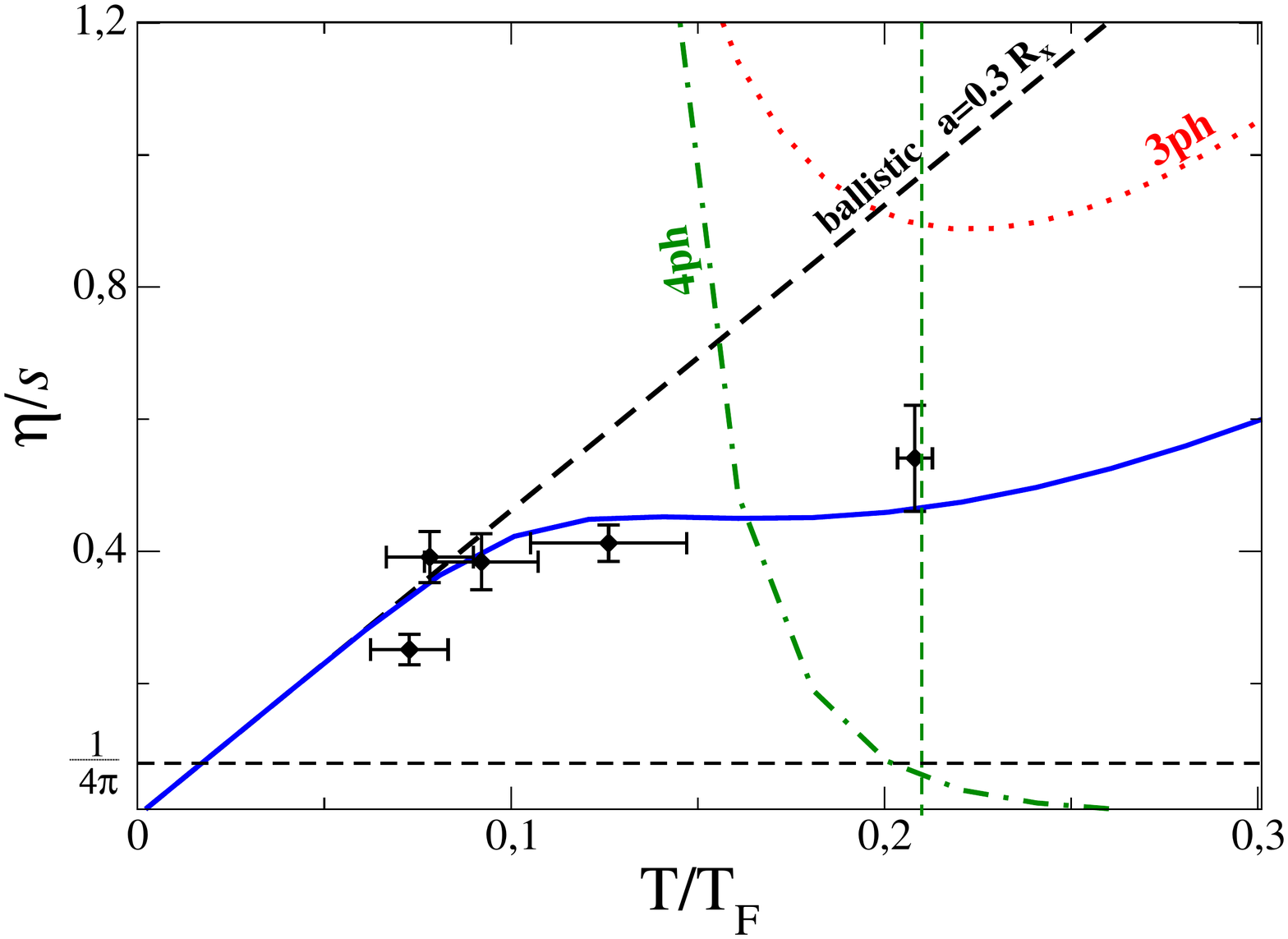}
\end{center}
\caption{Various contributions  to $\eta/s$ (in units of  $\hbar/k_B$) as a function of $T/T_F$. The  contribution of the interaction of the phonons with the boundary corresponds to the dashed black line.  The contribution of the 3ph process corresponding to the dotted  red line is obtained considering  $\tilde\delta \simeq 0.03$. The solid blue line corresponds to  $\eta_{\rm eff}$.  The vertical dashed green  line approximately corresponds to the phase transition temperature  \cite{Haussmann:2008}. The horizontal black dashed line corresponds to the universal bound derived in  \cite{hep-th/0104066}. The experimental values and error bars were taken from \cite{Cao:2010wa}. Adapted from \cite{Mannarelli:2012eg}. \label{fig:eta-s-total}}
\end{figure}

In Fig.\ref{fig:eta-s-total} we report several  contributions to the shear viscosity coefficient to entropy ratio. The interaction of the phonons with the boundary seems to play an important role for temperatures $T \lesssim 0.1 T_F$. Unfortunately few experimental data are available for such low temperature and it is hard to say whether the effective shear viscosity is able to describe the behavior of the system.

In the reported analysis a number of simplifying assumptions where used: 1) The shear viscosity is dominated by the center of the trap; the density in this region is bigger and roughly constant, which assures that  performing a  trap average might not modify drastically the results.
2) The fermionic contribution to $\eta$ was neglected, both in the superfluid core and in the outer normal layer; this should be a valid approximation at sufficiently low $T$, where superfluid fermions are known to be exponentially suppressed, while the outer fermionic layer of the cloud might be too dilute to provide enough damping. 3) The superfluid-normal fluid interface was modeled as a simple boundary, where phonons are specular-reflected or diffused according with the Maxwell-type condition; this approximation should be fine because  the Maxwell model is  appropriate for the description of an interface with no particle transfer and superfluid phonons are confined in the superfluid trap core. However, the microscopic description of the phonon interaction with unpaired fermions  in an optical trap has never been discussed and would certainly help to improve the (admittedly rough) Maxwell model. 

What is interesting here, is that a couple of testable predictions can be made from this study. First, we 
notice that if experiments are conducted reducing/increasing the size of the trap, but keeping $T_F$ constant, then $\eta/s$  at low temperatures should decrease/increase. In other words, the value of $\eta/s$ should correlate with the size of the
gas cloud. Second, we predict that $\eta/s$  should decrease with decreasing temperature. If we naively
extrapolate our results to lower temperature, we predict that there should be a
violation of the string theory proposed bound of $\eta/s$. Note, however, that this violation happens because phonons are  ballistic, while the string theory bound concerns the  hydrodynamic regime. 

Both the above-mentioned  predictions are independent of the detailed form of the phonon dispersion law, in particular, they are independent of the sign of the $\gamma$ term in Eq~\eqref{eq:disp-law} (which apparently is still a matter of debate).

\section{Conclusions}
The transport properties of a fluid in restricted geometries depend on the type of scattering that takes place at the boundary and on the geometrical properties of the boundary. We have evaluated the shear viscosity coefficient for a two-dimensional ballistic  phonon gas restricted by a circular boundary. If the boundary is put in rotation and if the phonons are diffused by the boundary, then a shear stress is exerted by the phonons on the boundary in a transient period. When phonons equilibrate with the boundary no shear stress is exerted on the boundary. These results have been applied to the case of phonons in an optical trap, where phonons live in the superfluid core and interact with the normal fluid of the corona.  Treating the interface   as a wall one has that if the shear viscosity is dominated by phonons, then $\eta/s$  should be proportional to the size of the system and to the temperature. \\

\noindent
{\bf Acknowledgments}\\
Part of the work presented here was studied with C.~Manuel and L.~Tolos  in \cite{Mannarelli:2012su, Mannarelli:2012eg}. I thank M.~Alford for discussion and  J.E.~Thomas and C.~Cao for providing their experimental data  of the shear to entropy ratio.


\begin{thebibliography}{99}
%\cite{Mannarelli:2012su, Mannarelli:2012eg}

%\cite{Mannarelli:2012su}
\bibitem{Mannarelli:2012su} 
  M.~Mannarelli, C.~Manuel and L.~Tolos,
  %``Shear viscosity in a superfluid cold Fermi gas at unitarity,''
  arXiv:1201.4006 [cond-mat.quant-gas].
  %%CITATION = ARXIV:1201.4006;%%

%\cite{Mannarelli:2012eg}
\bibitem{Mannarelli:2012eg} 
  M.~Mannarelli, C.~Manuel and L.~Tolos,
  %``Phonon contribution to the shear viscosity of a superfluid Fermi gas in the unitarity limit,''
  arXiv:1212.5152 [cond-mat.quant-gas].
  %%CITATION = ARXIV:1212.5152;%%


%----------------------- books rarefied gases -----------------------------------
\bibitem{Cercignani}
C.~Cercignani, \emph{ Rarefied gas dynamics}, Cambridge University Press, Cambridge, 2000
\bibitem{Sone}
Y.~Sone, \emph{ Kinetic theory and fluid dynamics}, Birkhauser, Boston, 2002
\bibitem{Struchtrup} H.~Struchtrup, 
 \emph{ Macroscopic transport equations for rarefied gas flows}, Springer, New York, 2005

%---------applications-------------

%\cite{Andersson:2010sh}
\bibitem{Andersson:2010sh} 
  N.~Andersson, B.~Haskell and G.~L.~Comer,
  %``r-modes in low temperature colour-flavour-locked superconducting quark star,''
  Phys.\ Rev.\ D {\bf 82}, 023007 (2010).
%  [arXiv:1005.1163 [astro-ph.SR]].
  %%CITATION = ARXIV:1005.1163;%%


%\cite{Manuel:2012rd}
\bibitem{Manuel:2012rd} 
  C.~Manuel and L.~Tolos,
  %``Shear viscosity and the r-mode instability window in superfluid neutron stars,''
  arXiv:1212.2075 [astro-ph.SR].
  %%CITATION = ARXIV:1212.2075;%%

\bibitem{Zadorozhko} A.~A.~Zadorozhko,  
{\'E}.~Y.~Rudavski{\u i},  V.~K.~Chagovets,  G.~A.~Sheshin,  
and  Y.~A.~Kitsenko, Low Temperature Physics, {\bf 35},  100 (2009).  

%%--------------review-------------
\bibitem{Giorgini:2008zz}
  S.~Giorgini, L.~P.~Pitaevskii and S.~Stringari,
  %``Theory of ultracold atomic Fermi gases,''
  Rev.\ Mod.\ Phys.\  {\bf 80}, 1215 (2008).
  %%CITATION = RMPHA,80,1215;%%


%%---------------Zwerger---------------
\bibitem{Haussmann:2008} 
R.~Haussmann and W.~Zwerger,
Phys.\ Rev.\ A {\bf 78}, 063602 (2008).
%[arXiv:0805.3226v4 [cond-mat.stat-mech]]. 


%---------MIT-------------

\bibitem{Ku}
M. J. H.~Ku, A. T.~Sommer, L. W.~Cheuk and M. W.~Zwierlein, 
Science {\bf 335}, 563 (2012).



%---------------------------c_1 and c_2------------------------
%\cite{Rupak:2008xq}
\bibitem{Rupak:2008xq}
  G.~Rupak and T.~Schafer,
  %``Density Functional Theory for non-relativistic Fermions in the Unitarity Limit,''
  Nucl.\ Phys.\ A\ {\bf 816}, 52  (2009).
%  [arXiv:0804.2678 [nucl-th]].
  %%CITATION = ARXIV:0804.2678;%%

\bibitem{Salasnich}
  L.~Salasnich and F.~Toigo,
  %``Viscosity-entropy ratio of the unitary Fermi gas from zero-temperature elementary excitations,''
  J.\ Low.\ Temp.\ Phys.\  {\bf 165}, 239 (2011).
%  [arXiv:1107.4552 [cond-mat.quant-gas]].
  %%CITATION = ARXIV:1107.4552;%%
  
  \bibitem{valle}
  J.~L.~Ma\~nes and M.~A.~Valle,
Annals of Physics {\bf 324}, 1136 (2009).% [arXiv:0810.3797 [cond-mat.other]].



%%---------------------------experiments Duke------------------------


\bibitem{kinast1}  %paper with the data used by us
J.~Kinast, A.~Turlapov and J. E.~Thomas, Phys.\ Rev.\ Lett.\  {\bf 94}, 170404 (2005).% [arXiv:cond-mat/0502507].
\bibitem{kinast3}
C.~Cao, E.~Elliott, H.~Wu and  J.E.~Thomas, NJP {\bf 13}, 075007 (2011).


%---------book-------------

\bibitem{Ashcroft-Mermin}
N.W.~Ashcroft, D.~Mermin, \emph{Solid State Physics}, Harcourt College Publisher, USA, 1976. 


%--------------------------AdS/CFT------------------------

%\cite{hep-th/0104066}
\bibitem{hep-th/0104066} 
  G.~Policastro, D.~T.~Son and A.~O.~Starinets,
  %``The Shear viscosity of strongly coupled N=4 supersymmetric Yang-Mills plasma,''
  Phys.\ Rev.\ Lett.\ \ {\bf 87}, 081601  (2001).
%  [hep-th/0104066].
  %%CITATION = PRLTA,87,081601;%%


\bibitem{Cao:2010wa} 
  C.~Cao, E.~Elliott, J.~Joseph, H.~Wu, J.~Petricka, T.~Schafer and J.~E.~Thomas,
  %``Universal Quantum Viscosity in a Unitary Fermi Gas,''
  Science {\bf 331}, 58 (2011).
%  [arXiv:1007.2625 [cond-mat.quant-gas]].
  %%CITATION = ARXIV:1007.2625;%%




\end{thebibliography}
\end{document}